\documentclass[sigconf]{acmart}

\usepackage{graphicx} 
\usepackage{float}
\usepackage{listings}
\usepackage{xcolor}
\usepackage{mdframed}
\usepackage[most]{tcolorbox}

\usepackage{amsmath,amsfonts}
\usepackage{textcomp}
\usepackage{array,booktabs}
\usepackage{multirow}
\usepackage{stfloats}
\usepackage{makecell}
\usepackage{colortbl}
\usepackage{algorithm}
\usepackage[noend]{algpseudocode}
\usepackage{balance}
\usepackage{verbatim}
\usepackage{tabularx}
\usepackage{xspace}
\usepackage{soul}
\usepackage{xurl}
\usepackage{hyperref}

\usepackage{titlesec}
\titlespacing*{\paragraph} {10pt}{0ex}{1ex}

\definecolor{cellh}{gray}{1.0} 

\AtBeginDocument{%
  }

\acmConference[Preprint]{}{2026}{Raleigh, NC, USA}

\newcommand{\numtrajectories}{9,374\xspace}

\newcommand{\numagents}{19\xspace}
\newcommand{\numagentframeworks}{8\xspace}
\newcommand{\numpythonframeworks}{12\xspace}
\newcommand{\numllms}{14\xspace}
\newcommand{\numtasks}{500\xspace}
\newcommand{\numtaskfeatures}{90\xspace}

\newcommand{\numfeatpatch}{12\xspace}
\newcommand{\numfeattest}{15\xspace}
\newcommand{\numfeatissue}{53\xspace}
\newcommand{\numfeatmeta}{10\xspace}

\setcopyright{none}
\renewcommand\footnotetextcopyrightpermission[1]{}

\begin{document}


\title[Behavioral Drivers of Coding Agent Success and Failure]{Beyond Resolution Rates: Behavioral Drivers of Coding Agent Success and Failure}

\author{Tural Mehtiyev}
\affiliation{%
  \institution{North Carolina State University}
  \city{Raleigh}
  \state{NC}
  \country{USA}
}

\author{Wesley Assunção}
\affiliation{%
  \institution{North Carolina State University}
  \city{Raleigh}
  \state{NC}
  \country{USA}
}

\begin{abstract}
Coding agents represent a new paradigm in automated software engineering, combining the reasoning capabilities of Large Language Models (LLMs) with tool-augmented interaction loops. However, coding agents still have severe limitations. Top-ranked LLM-based coding agents still fail on over 20\% of benchmarked problems. Yet, we lack a systematic understanding of \emph{why} (i.e., the causes) agents fail, and \emph{how} failure unfolds behaviorally. We present a large-scale empirical study analyzing \numtrajectories trajectories from \numagents agents (\numagentframeworks coding agent frameworks, \numllms LLMs) on \numtasks tasks. 
We organize our analysis around three research questions. First, we investigate why agents fail on specific tasks and find that patch complexity alone does not explain difficulty: 12 never-solved tasks require only simple patches and were considered easy by human annotators, yet all agents fail due to gaps in architectural reasoning and domain knowledge. Second, we examine how behavioral patterns differentiate success from failure. The widely reported correlation between trajectory length and failure reverses direction once task difficulty is controlled, revealing it as a confound. Instead, trajectory \emph{structure} discriminates consistently: agents that gather context before editing and invest in validation succeed more often, and these strategies are agent-determined rather than task-adaptive. Third, we disentangle LLM capability from framework design and find that the LLM is the primary driver of both outcome and behavior: agents sharing the same LLM agree on far more tasks than agents sharing the same framework, and the framework performance gap shrinks with each generation of LLM improvement. Framework prompts do influence agent tactics, but this influence diminishes with stronger LLMs.
\end{abstract}

\maketitle

\section{Introduction}
\label{sec:introduction}

Coding agents represent a new paradigm in automated software engineering, combining the reasoning capabilities of Large Language Model (LLM) with tool-augmented interaction loops that allow them to operate directly on codebases~\cite{liu2024llmagents_se_survey}. Unlike traditional code generation systems that produce isolated snippets~\cite{chen2021codex,jiang2024codegen_survey}, coding agents function as autonomous developers~\cite{yang2024sweagent,xia2025agentless}: they interpret natural-language issue reports, explore repositories, execute code, run tests, and iteratively refine patches to resolve a task. Frameworks such as SWE-agent~\cite{yang2024sweagent}, OpenHands~\cite{wang2024openhands}, and AutoCodeRover~\cite{zhang2024autocoderover} have demonstrated the feasibility of this approach, leading to increasing adoption in real-world workflows. Ehsani et al.~\cite{ehsani2026where} report over 33k agent-authored pull requests on GitHub, and the SWE-bench leaderboard~\cite{jimenez2024swebench} tracks over 80 unique approaches across dozens of competing systems~\cite{martinez2025dissecting}. 
Despite this rapid progress and growing ecosystem, coding agents remain far from reliable. State-of-the-art agents still fail on more than 20\% of SWE-bench Verified tasks~\cite{swebench_leaderboard} (as of February 2026), and failure rates increase sharply on harder benchmarks~\cite{swebenchpro2025}, highlighting the need for a deeper understanding of their behavior.

To contribute to the construction of reliable agentic systems, understanding \textit{why} and \textit {how} coding agents fail is therefore critical. The ``why'' refers to the underlying causes of failure, namely whether failures arise from task characteristics (e.g., hidden dependencies, test requirements), limitations of the LLM (e.g., reasoning errors, poor generalization), or constraints imposed by the agent framework (e.g., tool orchestration, context management). Additionally, the ``how'' refers to the observable behavioral process through which failure unfolds, primarily captured in the agent's trajectory as a sequence of actions such as exploration, editing, testing, and iteration~\cite{yao2023react}. While prior work has studied outcome-level metrics such as resolution rates and fault localization accuracy~\cite{meng2024bugfixing}, classified failure root causes through manual inspection~\cite{empiricalfailures2025}, or analyzed behavioral patterns such as trajectory length and action sequences in small agent samples~\cite{majgaonkar2025,bouzenia2025understanding,chen2026beyondfinalcode,liu2026graphectory}, these studies rarely connect the \textit{why} and \textit{how} dimensions of failure while controlling for task difficulty on a large-scale dataset. As a consequence, there is a lack of systematic understanding of how specific behaviors lead to particular failure causes, and which aspects of agent design or capability should be improved. This gap limits both scientific insight and practical progress, as improving agents requires not just knowing that they fail, but understanding the mechanisms by which those failures occur.


In this paper, we address this gap by conducting a large-scale empirical study of coding agent behavior. We analyze \numtrajectories trajectories from 19 agents spanning \numagentframeworks frameworks and \numllms LLMs, all evaluated on the same 500 SWE-bench Verified tasks. In the studied dataset, every agent attempts every task, enabling within-task comparisons that control for task difficulty, an important confound that has undermined prior work.
Our goal is to move beyond outcome-level benchmarking and instead understand the behavioral drivers of success and failure. We focus on three core research questions:
\textbf{\textit{RQ1. Why do coding agents fail on certain tasks?}}
Prior work mainly approximates difficulty through proxy metrics such as patch size~\cite{swebenchpro2025} or human-estimated solve time~\cite{ganhotra2025cracking,zan2025multiswebench}. We test whether this holds across all 500 tasks and investigate edge cases in which some tasks are considered ``easy'' yet every agent fails.
\textbf{\textit{RQ2. How do the behavioral patterns differentiate success from failure?}} We use within-task paired comparisons to test whether the widely-reported length--failure correlation survives difficulty controls, and characterize how different agents fail on the same tasks.
\textbf{\textit{RQ3. Does LLM capability or framework design drive agent success?}} By leveraging natural variation in LLMs within the same frameworks, we disentangle the relative contributions of model capability and system architecture.

\begin{figure*}[t]
    \centering
    \includegraphics[width=.95\textwidth]{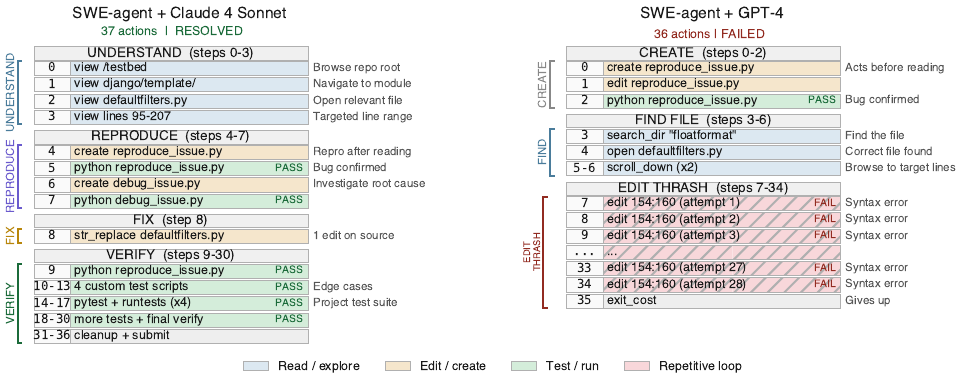}
    \caption{Contrasting agent trajectories on django-15863. Both agents use the SWE-agent framework. Claude~4~Sonnet (left) follows a structured workflow: browse the codebase, reproduce the bug, make one surgical edit, then run several verification cycles before submitting. GPT-4 (right) finds the correct file but then edits the code many times, producing 28 syntax errors, without ever running the project's test suite.}
    \label{fig:django15863}
\end{figure*}

Through both quantitative and qualitative analysis, our large-scale, controlled study links coding agent behavior to outcomes across diverse systems. The results lead to several key findings:
(i) simple-patch tasks can defeat every agent when they require architectural reasoning rather than complex edits;
(ii) trajectory length is an ambiguous failure signal: its direction reverses depending on whether one controls for agent identity or task difficulty;
(iii) trajectory structure distinguishes success from failure. Agents that gather context before editing and validate more succeed consistently, and these strategies are fixed across task complexity; and 
(iv)  LLM capability dominates over framework design. As LLMs improve, framework differences shrink and agents converge. 

\section{Background and Motivation}
\label{sec:background}

Agents are built on top of a \textit{framework} (also referred to as scaffold in this paper; e.g., SWE-agent~\cite{yang2024sweagent} and OpenHands~\cite{wang2024openhands}) that orchestrates tool use, manages context, and structures the problem-solving loop, wrapped around an \textit{LLM} that generates each action in the trajectory~\cite{wang2024llmagent_survey}. To solve software development tasks, agents produce a \emph{trajectory}, a step-by-step sequence of actions (file reads, edits, test executions, search queries) and the environment's responses (compiler output, test results, error messages). Trajectories are the behavioral equivalent of a developer's actions, capturing not just whether the agent succeeded, but \emph{how} it approached the problem. This behavioral record is the primary data source for understanding agent strategies, failure modes, and intervention opportunities~\cite{majgaonkar2025,bouzenia2025understanding}.

Coding agents show promising results on several benchmarks~\cite{jimenez2024swebench,swebench_leaderboard}, but they still present significant limitations. As of February 2026, the best submission on the SWE-bench Verified leaderboard~\cite{swebench_leaderboard} fails on more than 20\%, meaning that even the strongest agent fails on nearly one in five tasks. 
When an agent fails, it wastes computing resources, developer review time, and organizational trust. In the SWE-bench Verified dataset we study, 12.4\% of total compute is spent on the 55 tasks that no agent solves, resources with zero return. Thus, it is paramount to understand why and how coding agents fail.

To illustrate what can influence an agent's success, we focus on the task django-15863 in SWE-bench Verified~\cite{jimenez2024swebench,swebenchverified2024}: a 1-line fix that causes \texttt{floatformat} to drop precision on \texttt{Decimal} numbers. Three agents in the dataset failed to solve this task. 
This indicates that the \textit{task} may be too complex for the same agents (\textbf{RQ1}: Task difficulty). 
Figure~\ref{fig:django15863} shows trajectories for two agents working on the task django-15863, one with Claude~4~Sonnet (on the left) and one with GPT-4 (on the right), but both using the same framework.
The two agents produced similar numbers of actions in their trajectories: 37 for Claude~4~Sonnet and 36 for GPT-4.
The agent with Claude~4~Sonnet makes one edit on the source code, has zero syntax errors, and solves the task successfully. GPT-4 makes many edits, with 28 syntax errors, and consequently fails. 
We observe in this example that, despite having similar numbers of steps in their trajectories, the actions differ significantly. The trajectory of the agent with Claude~4~Sonnet follows a \texttt{``understand > reproduce > fix > verify''} workflow, whereas the agent with GPT-4 follows \texttt{``create > find > edit''}. Thus, we conjecture that the success or failure of an agent may be related to the patterns of actions in its trajectory (\textbf{RQ2}: behavioral patterns). Finally, agents rely on a \textit{Framework} and have an \textit{LLM}, which produces the trajectories. Therefore, we hypothesize that both LLM and framework, and their interactions with the tasks, may contribute to success or failure in ways that existing work has not explored (\textbf{RQ3}: LLM capability and framework design). 

\section{Study Design}
\label{sec:methodology}

This section describes the data collection, trajectory encoding, feature-extraction pipeline, and analytical methods to answer the three RQs of our study.

\subsection{Data Collection}
\label{sec:data-collection}

Our study relies on \numtrajectories trajectories from \numagents agents on the \numtasks SWE-bench Verified tasks~\cite{jimenez2024swebench,swebenchverified2024}, which target real-world bugs across \numpythonframeworks open-source Python repositories. Agents represent \numagentframeworks coding agent frameworks: SWE-agent, OpenHands, Trae, Skywork, CodeSweep, EPAM-AI, SAGE, and Sonar. Additionally, the dataset represents \numllms LLMs: four Claude variants, three GPT variants, and seven others (e.g., Kimi-K2, Devstral, Qwen-32B, and Doubao). All trajectories are collected from submissions to the SWE-bench Verified leaderboard~\cite{swebench_leaderboard}, downloaded from the public experiment bucket maintained by the benchmark authors~\cite{jimenez2024swebench}. Each agent attempts all \numtasks tasks, yielding a complete  $agent \times task$  matrix, with minor gaps: three agents have 443--499 runs due to timeouts or infrastructure failures. 
Since every task is attempted by every agent, this enables within-task comparisons that control for task difficulty, a confound that prior studies with 1--3 agents could not address.

\subsection{Trajectory Encoding}
\label{sec:feature-extraction}

Previous work classifies agent trajectories at varying granularity: Bouzenia and Pradel~\cite{bouzenia2025understanding} define eight action types (Explore, Locate, Search, Reproduce, Generate fix, Run tests, Refactor, Explain), while Graphectory~\cite{liu2026graphectory} abstracts these into four steps (Localization, Patching, Validation, General). Both focus solely on the agent's actions. We expand this by jointly encoding the agent action and the environment's response. For example, a clean patch (i.e., edit) is encoded as P, while an edit that triggers a SyntaxError is encoded as Ps, and a test run that passes is Vp (i.e., Validation pass), while one that fails is Vf (i.e., Validation fail). 
Our encoding, shown in Table~\ref{tab:enriched-symbols}, comprises 13 sub-step symbols that capture execution quality and error type information while remaining compact and human-readable.

\begin{table}[t]
\caption{Enriched encoding: 13 sub-phase symbols. Each symbol is classified from the raw \texttt{.traj} data using deterministic regex rules on the action and observation fields.}
\label{tab:enriched-symbols}
\centering
\small
\addtolength{\tabcolsep}{-3pt}
\begin{tabular}{llp{3.8cm}}
\toprule
\textbf{Symbol} & \textbf{Meaning} & \textbf{Source in trajectory} \\
\midrule
Lb & Locate $\to$ browse whole file & \texttt{view} w/o \texttt{--view\_range} \\
Lt & Locate $\to$ targeted read & \texttt{view} with \texttt{--view\_range}, \texttt{goto} \\
Ls & Locate $\to$ with search & \texttt{grep}, \texttt{search\_dir}, \texttt{search\_file} \\
\midrule
P  & Patch (clean edit) & edit with no error in observation \\
Ps & Patch $\to$ syntax error & edit followed by SyntaxError \\
Pi & Patch $\to$ import error & edit followed by ImportError \\
Pr & Re-patch (same file) & edit on previously edited file \\
\midrule
Vp & Validation passed & test run with no error \\
Vf & Validation $\to$ test failure & test run with test failure \\
Ve & Validation $\to$ runtime error & test run with RuntimeError \\
Vr & Reproduction script & \texttt{reproduce\_*.py} or similar \\
\midrule
E  & Environment setup & pip, apt, conda, build \\
G  & General / other & submit, cd, pwd, think \\
\bottomrule
\end{tabular}
\end{table}

\subsection{Task-Level Feature Extraction}
\label{sec:feature-extraction-tasks}

For each of the 500 tasks, we extract \numtaskfeatures features. Table~\ref{tab:task-features} summarizes the four categories and representative features. Our feature selection draws on established dimensions from prior work: (i) \emph{Patch complexity} features follow the metrics used as difficulty proxies in prior SWE-bench studies~\cite{zan2025multiswebench,ganhotra2025cracking,swebenchpro2025}; (ii) \emph{Test demand} features capture how demanding the task's test infrastructure is, using the fail-to-pass test structure defined by SWE-bench~\cite{jimenez2024swebench}; (iii) \emph{Issue/prompt} features are informed by the bug report quality literature, which shows that steps to reproduce, stack traces, and code examples are the most impactful elements for bug resolution~\cite{bettenburg2008bugreport,weiss2007fixtime}; and (iv) \emph{Metadata} features include repository, version, and issue age to control for project-level variation, as cross-project studies show that different projects have fundamentally different difficulty profiles~\cite{zimmermann2009crossproject}.
Features are computed deterministically using regex-based parsers on the issue text and diff-based analysis on the patches. The full list of features is in our supplementary material~\cite{supplementar}.


\begin{table}[t]
\caption{Task-level features (\numtaskfeatures total) extracted per task, organized by source.}
\label{tab:task-features}
\centering

\addtolength{\tabcolsep}{-3pt}
\small
\begin{tabular}{lrp{5cm}}
\toprule
\textbf{Category} & \textbf{Count} & \textbf{Representative features} \\
\midrule
Patch complexity & \numfeatpatch & lines added/deleted, hunks, files modified, multi-file flag, hunk size, directory depth \\
Test demand & \numfeattest & fail-to-pass count, test assertions, parametrized tests, nesting depth \\
Issue / prompt & \numfeatissue & prompt length (words, sentences), has stacktrace, has repro code, code-to-text ratio, specificity score, lexical diversity \\
Metadata & \numfeatmeta & repository, version, issue age, repo popularity, domain \\
\bottomrule
\end{tabular}
\end{table}

  \subsection{Analytical Design}
 \label{sec:paired-comparison}

\subsubsection{RQ1: Task difficulty and agent resolution analysis.}
To answer RQ1, we conducted a quantitative and qualitative analysis of tasks considered ``easy'', yet all agents failed to solve. We isolate 12 never-solved tasks that have simple patches (single file, $\leq$10 changes) and compare them against 25 always-solved tasks with similar patch characteristics using Mann-Whitney $U$ tests and Cliff's $\delta$ across \numtaskfeatures task-level features. Where quantitative features do not fully explain the gap, we turn to qualitative trajectory analysis: for each of the 12 tasks, we examine the trajectories by the best-performing agents and compare their submitted patch against the gold patch to identify the failure mode.

\subsubsection{RQ2: Paired comparison of agent trajectories.}
Comparing resolved and failed trajectories requires controlling for two confounds simultaneously: \textit{agent identity}, since different agents have different capabilities; and \textit{task difficulty}, since harder tasks produce more failures. Thus, we design the analysis around two approaches, formalized in Algorithm~\ref{alg:rq2}. \emph{Approach~A} fixes the agent and compares its resolved vs.\ failed trajectories across tasks, controlling for agent identity but confounded by task difficulty. \emph{Approach~B} fixes the task and compares resolved vs.\ failed trajectories from different agents, controlling for task difficulty but confounded by agent identity. 

We apply this analytical design to answer RQ2, using trajectory length as the metric $m$ for RQ2a and per-symbol enriched encoding frequency (e.g., proportion of steps spent on validation) for RQ2b. We set $n_{\min} = 5$ to ensure stable estimates.
Prior studies~\cite{majgaonkar2025,bouzenia2025understanding} compare successes and failures without this control, so their findings are confounded by task difficulty. Our complete agent$\times$task matrix makes both within-agent and within-task comparisons possible.

\subsubsection{RQ3: Agent framework vs. LLM capability.}

We exploit a natural experiment: by holding the framework constant and varying the LLM, we isolate the LLM effect; by holding the LLM constant and varying the framework, we isolate the framework effect. We compare both factors using aggregate resolution rates and per-task outcome agreement. Finally, we analyze the prompt's effect on overall agent strategy.

\begin{algorithm}[t]
\caption{Two-Approach Success/Failure Comparison}
\label{alg:rq2}
\begin{algorithmic}[1]
\Require Trajectories $\mathcal{T}$, contested tasks $\mathcal{C}$ (tasks with both resolved and failed outcomes), behavioral metric $m$, min.\ sample size $n_{\min}$, significance level $\alpha$
\Ensure Per-approach effect sizes and significance
\Statex \hspace{1em} A positive $\delta$ or $\Delta$ indicates $m$ is higher for failed trajectories.
\Statex
\Statex \textbf{--- Approach A: fix agent, vary all tasks (one-sided) ---}
\For{each agent $a \in \text{Agents}$}
    \State $R_a \gets \{t \in \mathcal{T} : t.\text{agent} = a,\; t.\text{resolved}\}$
    \State $F_a \gets \{t \in \mathcal{T} : t.\text{agent} = a,\; t.\text{failed}\}$
    \If{$|R_a| < n_{\min}$ \textbf{or} $|F_a| < n_{\min}$} \textbf{skip} \EndIf
    \State $\delta_a, p_a \gets \text{Mann-Whitney } U \text{ (one-sided: } F > R\text{)} + \text{Cliff's } \delta$ on $m(F_a)$ vs $m(R_a)$
\EndFor
\State \textbf{Aggregate:} count agents with $\delta_a > 0$ ($p < \alpha$) vs $\delta_a < 0$ ($p < \alpha$)
\Statex
\Statex \textbf{--- Approach B: fix task, vary agents (two-sided) ---}
\For{each task $t \in \mathcal{C}$}
    \State $R_t \gets \{\tau \in \mathcal{T} : \tau.\text{task} = t,\; \tau.\text{resolved}\}$
    \State $F_t \gets \{\tau \in \mathcal{T} : \tau.\text{task} = t,\; \tau.\text{failed}\}$
    \State $\Delta(t) \gets \text{mean}(m(F_t)) - \text{mean}(m(R_t))$
\EndFor
\State $p \gets$ Wilcoxon signed-rank (two-sided) on $\{\Delta(t)\}_{t \in \mathcal{C}}$
\State \textbf{Report:} $\bar{\Delta}$, $p$, fraction of tasks with $\Delta(t) > 0$
\end{algorithmic}
\end{algorithm}

\section{Results and Discussion}
\label{sec:results}

This section presents the results and discussion organized per RQ.

\subsection{RQ1: Why do coding agents fail on certain tasks?}
\label{sec:results-rq1}

\subsubsection{Difficulty Landscape and Edge-Case Selection.}

\begin{figure}[t]
    \centering
    \includegraphics[width=\columnwidth]{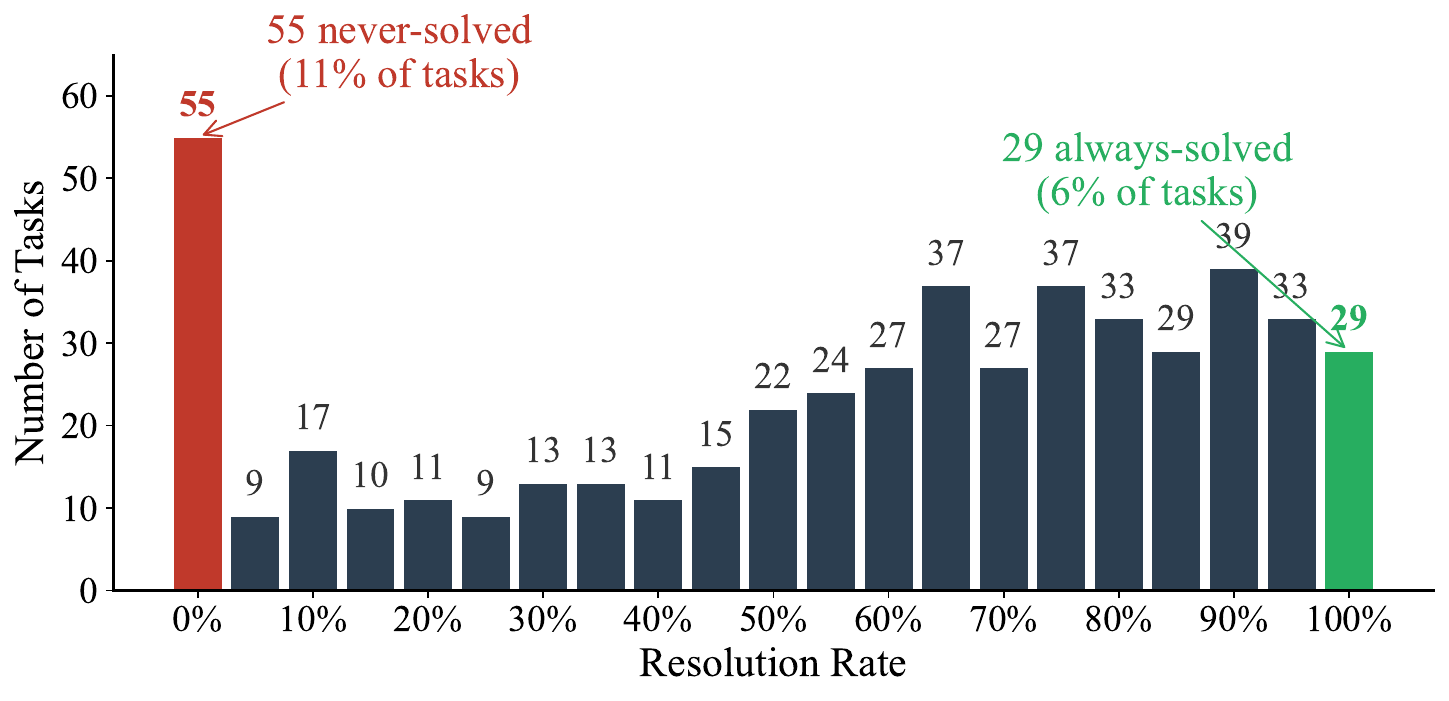}
    \caption{Task difficulty distribution across 500 SWE-bench Verified tasks.}
    \label{fig:rq1-difficulty}
\end{figure}

We classify each of the 500 tasks by its resolution rate across all agents. The distribution is bimodal (Figure~\ref{fig:rq1-difficulty}): 55 tasks (11\%) are never solved by any agent, 416 (83\%) are contested, and 29 (6\%) are always solved.

Among the 55 never-solved tasks, 12 require only a single-file patch of $\leq$10 total changes. Table~\ref{tab:simple-never-solved} presents these tasks. Focusing on patch complexity, as in prior work~\cite{zan2025multiswebench,ganhotra2025cracking,swebenchpro2025}, these are considered simple tasks by humans and are expected to be solved at high rates across our dataset.
Therefore, these edge cases invalidate the prior work's conclusions that correlate patch complexity with difficulty, suggesting that something beyond patch size must also matter.

To further investigate these cases, we compare the 12 never-solved tasks with the 25 always-solved tasks that have similarly simple patches (single file, $\leq$10 changes). The results show that patch complexity is statistically indistinguishable between the two groups (mean total changes: 4.75 vs.\ 3.76; with Mann-Whitney $p = 0.24$, Cliff's $\delta = 0.24$). Of the \numtaskfeatures task-level features we investigated,\footnote{The complete list of features is available in our supplementary material~\cite{supplementar}.} Table~\ref{tab:rq1-sig-features} presents the five ones with significant results ($p < 0.05$), all of which are test-demand or issue-description features. However, three of the twelve never-solved tasks have the minimum possible test demand (FTP\,=\,1), and when we restrict the comparison to very simple patches ($\leq$5 changes), the test-demand differences become non-significant. Quantitative features alone do not fully distinguish these edge cases, motivating a detailed manual qualitative analysis of the agent trajectories themselves.

\begin{table}[!tp]
\caption{Twelve never-solved tasks with simple patches. The \emph{Human label} column shows the official difficulty annotation from SWE-bench Verified's human annotators~\cite{swebenchverified2024}. By every prior difficulty metric and by human judgment, these are easy tasks. Zero of \numagents agents solve any of them.}
\label{tab:simple-never-solved}
\centering
\small
\begin{tabular}{lrrrrl}
\toprule
\textbf{Task} & \textbf{Lines} & \textbf{Hunks} & \textbf{Files} & \textbf{Res.} & \textbf{Human label} \\
\midrule
django-11477       & 2  & 1 & 1 & 0/19 & 15 min--1 hr \\
django-15098       & 2  & 1 & 1 & 0/19 & 15 min--1 hr \\
django-16667       & 2  & 1 & 1 & 0/19 & 15 min--1 hr \\
sympy-20428        & 2  & 1 & 1 & 0/19 & 15 min--1 hr \\
matplotlib-23476   & 3  & 1 & 1 & 0/19 & $<$15 min \\
matplotlib-23299   & 5  & 2 & 1 & 0/19 & 15 min--1 hr \\
django-13794       & 6  & 1 & 1 & 0/19 & $<$15 min \\
sympy-21930        & 6  & 3 & 1 & 0/19 & 15 min--1 hr \\
django-10999       & 7  & 1 & 1 & 0/19 & $<$15 min \\
django-14792       & 7  & 1 & 1 & 0/19 & $<$15 min \\
matplotlib-21568   & 7  & 1 & 1 & 0/19 & 15 min--1 hr \\
django-15252       & 8  & 1 & 1 & 0/19 & 15 min--1 hr \\
\bottomrule
\end{tabular}
\end{table}

\begin{table}[t]
\centering
\caption{The 5 features (of \numtaskfeatures tested) reaching $p < 0.05$ between 12 never-solved and 25 always-solved simple-patch tasks. No patch or metadata feature is significant.}
\label{tab:rq1-sig-features}
\small
\begin{tabular}{lrrrc}
\toprule
\textbf{Feature} & \textbf{Never} & \textbf{Always} & \textbf{Cliff's $\delta$} & \textbf{$p$} \\
\midrule
Fail-to-pass count & 2.42 & 1.24 & +0.61 (L) & $<$0.001 \\
FTP test files & 2.17 & 1.20 & +0.55 (L) & 0.001 \\
Test patch files & 1.58 & 1.00 & +0.33 (M) & 0.003 \\
Test patch changes & 21.25 & 12.64 & +0.46 (M) & 0.025 \\
Has repro code & 0.67 & 0.24 & +0.43 (M) & 0.014 \\
\bottomrule
\end{tabular}
\end{table}

\subsubsection{Trajectory Analysis: The Architectural Reasoning Gap.}

\begin{figure}[!tp]
    \centering
    \includegraphics[width=\columnwidth]{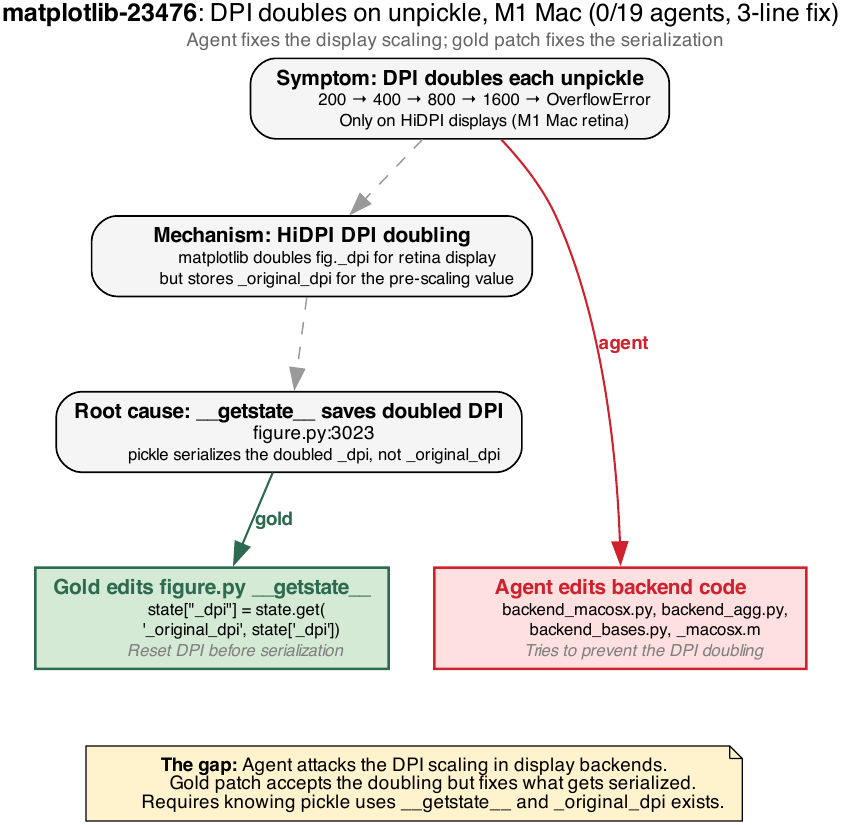}
    \caption{Representative case study: matplotlib-23476. The agent fixes DPI scaling in display backends (4 files); the gold patch fixes the serialization in \texttt{\_\_getstate\_\_} (3 lines). Both address the same bug from different architectural levels.}
    \label{fig:rq1-case-mpl23476}
\end{figure}

Figure~\ref{fig:rq1-case-mpl23476} illustrates what we mean by \emph{architectural reasoning gap} through the task matplotlib-23476. On HiDPI screens (e.g., M1 Mac), matplotlib internally doubles the figure DPI for retina display but tracks the original value in a separate attribute (\texttt{\_original\_dpi}). The bug is that \texttt{Figure.\_\_getstate\_\_()}, the method Python's pickle protocol calls during serialization, saves the doubled DPI instead of the original. Each pickle/unpickle cycle doubles the DPI (200$\to$400$\to$800$\to\ldots$). The 3-line gold patch resets \texttt{\_dpi} to \texttt{\_original\_dpi} before serialization. The best agent (61 steps) correctly finds \texttt{figure.py}, understands the DPI doubling, but tries to prevent the doubling from happening by editing four backend files (\texttt{backend\_macosx.py}, \texttt{backend\_agg.py}, \texttt{backend\_bases.py}, \texttt{\_macosx.m}). It attacks the display scaling layer instead of the serialization layer. Both the agent and the gold patch address the same symptom, but at different architectural levels: the agent tries to stop the value from changing, while the fix ensures the changed value is not persisted. The qualitative analysis for all 12 edge cases is available in our supplementary material~\cite{supplementar}.

The pattern observed in the example above is systematic. We examined the trajectories of the top-performing agents on all 12 never-solved simple-patch tasks (Table~\ref{tab:simple-never-solved}), comparing their submitted patches against the gold patch. Table~\ref{tab:architectural-gap} summarizes the results. In every case, the agent correctly localizes the bug, finding the gold-patch file in 12 of 12 tasks and editing it in 10 of 12. Yet it consistently intervenes at the wrong level. In 10 of 12 tasks, the failure traces back to architectural judgment: the agent patches the symptom (the caller, the consumer, the display layer) while the gold patch fixes the root cause (the callee, the producer, the serialization layer), or the agent cannot judge the right scope for its fix. The remaining two tasks involve a behavioral failure (abandoning a correct fix mid-trajectory) and a domain knowledge gap (TeX math-mode formatting). The agent's fixes compile and pass its own tests; they fail because the SWE-bench test suite validates behavior at the root-cause level.

\begin{table}[!tp]
\centering
\caption{Architectural gap across all 12 never-solved simple-patch tasks. The best agent found the gold-patch file in 12/12 and edited it in 10/12, yet failed every time.}
\label{tab:architectural-gap}
\small
\begin{tabular}{lll}
\toprule
\textbf{Task} & \textbf{Gap type} & \textbf{Root cause} \\
\midrule
django-11477 & Caller $\to$ Callee & \multirow{10}{*}{\parbox{2.4cm}{\emph{Architectural judgment} (10 tasks)}} \\
django-13794 & Symptom $\to$ Root cause & \\
django-14792 & Consumer $\to$ Producer & \\
mpl-23299 & Add $\to$ Remove behavior & \\
mpl-23476 & Display $\to$ Serialization & \\
sym-20428 & Operations $\to$ Type system & \\
django-15252 & Components $\to$ Orchestrator & \\
django-10999 & Issue's fix $\to$ Redesign & \\
sym-21930 & Local + general $\to$ Local only & \\
django-16667 & Wrong control flow & \\
\midrule
django-15098 & Abandoned correct fix & \emph{Behavioral} \\
mpl-21568 & Wrong TeX commands & \emph{Domain knowledge} \\
\bottomrule
\end{tabular}
\end{table}

\begin{tcolorbox}[colback=gray!5, colframe=black, title=\textbf{Answering RQ1}, breakable]
Patch complexity alone does not explain why agents fail. Twelve never-solved tasks require simple patches ($\leq$10 changes, single file) and were labeled easy by human annotators. Quantitative analysis shows test demand and issue-description features distinguish these from always-solved tasks, but the deeper cause is an \emph{architectural reasoning gap}: agents correctly localize the bug and produce plausible fixes, but intervene at the wrong architectural layer (10/12 tasks). They patch symptoms rather than root causes.
\end{tcolorbox}

\paragraph{Implications.} RQ1 reveals a ceiling that even top-performing agents may not overcome. The 12 never-solved simple-patch tasks fail not because agents lack the right tool or take too few steps, but because they intervene at the wrong architectural layer. Agents find the correct file and produce plausible fixes, yet consistently patch the symptom rather than the root cause. This architectural reasoning gap suggests that scaling model size or training on more code may not suffice; agents may need explicit mechanisms for reasoning about component boundaries, ownership, and the direction of causal dependencies in a codebase. For benchmark designers, this implies that test-demand metrics and architectural complexity indicators should complement patch size as difficulty measures.

\subsection{RQ2: How do the behavioral patterns differentiate success from failure?}
\label{sec:results-rq2}

\subsubsection{RQ2a: Trajectory Length Shows a Confounding Reversal}
\label{sec:results-length-failure}

Prior work reports that failed agent trajectories are substantially longer than successful ones. Majgaonkar et al.~\cite{majgaonkar2025} find 12.6--82.5\% longer failed trajectories across agents on SWE-bench. However, these comparisons pool trajectories across tasks of varying difficulty without controlling for it. We apply two approaches (see Section~\ref{sec:paired-comparison}) to investigate whether this finding holds.

\begin{figure}[t]
\centering
\includegraphics[width=\columnwidth]{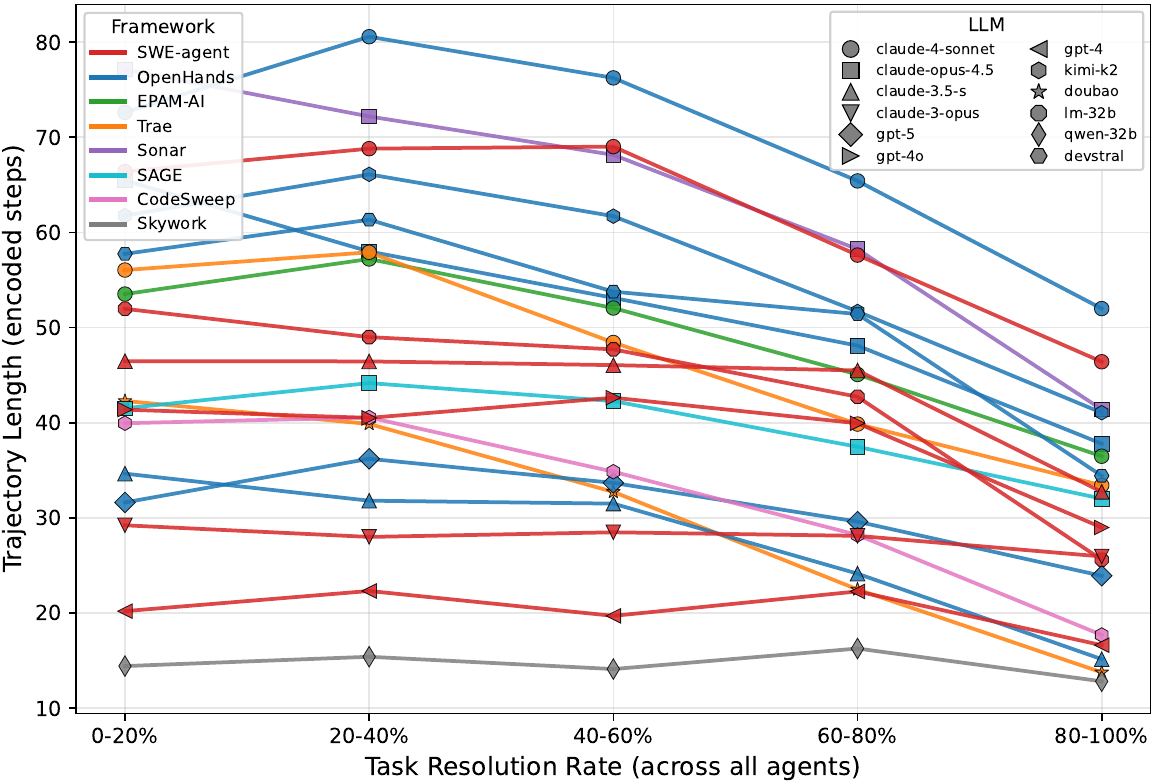}
\caption{Trajectory length by task difficulty for all 19~agents. Every agent takes more steps on harder tasks (all Spearman $\rho < 0$, all $p < 0.01$). This universal length--difficulty coupling explains the confounding reversal: length reflects task difficulty, not strategy quality.}
\label{fig:rq2a-length-difficulty}
\end{figure}

\paragraph{Approach~A (fix agent, vary tasks).} We replicate the cross-task finding across all agents on SWE-bench Verified. Table~\ref{tab:length-failure-all} shows that for every agent, failed trajectories are significantly longer ($p < 0.001$, one-sided Mann-Whitney $U$), with percentage gaps from $+$14.0\% to $+$111.9\%. The effect is universal: regardless of framework or LLM, when a given agent fails, it takes more steps than when it succeeds.
However, Approach~A is confounded by task difficulty. Figure~\ref{fig:rq2a-length-difficulty} shows the source: for every agent, trajectory length increases with task difficulty ($\rho$ from $-$0.13 to $-$0.57, all $p < 0.01$). Tasks with a lower resolution rate produce longer trajectories regardless of outcome, so the within-agent ``failures are longer'' effect may partly reflect that agents fail on harder (and therefore longer) tasks.

\begin{table}[!tp]
\caption{Length--failure effect. \textbf{Approach~A} (top): for each agent, mean steps of resolved vs.\ failed runs across all tasks; every agent shows failures are significantly longer ($p < 0.001$, one-sided Mann-Whitney $U$). \textbf{Approach~B} (bottom): for each contested task, mean steps of resolved vs.\ failed agents; the effect \emph{reverses}: resolved trajectories are longer on 63\% of tasks ($p < 10^{-8}$, two-sided Wilcoxon signed-rank).}
\label{tab:length-failure-all}
\centering
\addtolength{\tabcolsep}{-3pt}
\small
\begin{tabular}{l|l|r|r|r|r}
\toprule
\textbf{Agent} & \textbf{LLM} & \textbf{Res.\,Rate} & \textbf{\% Longer} & \textbf{Cliff's $\delta$} & \textbf{Sig.} \\
\midrule
\multicolumn{6}{c}{\emph{Approach~A: fix agent, vary all tasks (one-sided Mann-Whitney $U$)}} \\
\midrule
Sonar      & claude-opus-4.5       & 79.2\% & +30.5\% & +0.434 & *** \\
Trae       & doubao-seed-code      & 78.5\% & +65.1\% & +0.316 & *** \\
OpenHands  & claude-opus-4.5       & 78.3\% & +29.6\% & +0.403 & *** \\
EPAM-AI    & claude-4-sonnet       & 76.8\% & +19.9\% & +0.389 & *** \\
Trae       & claude-4-sonnet+opus  & 75.4\% & +36.7\% & +0.503 & *** \\
SAGE       & claude-4.5+gpt-5      & 73.0\% & +14.5\% & +0.224 & *** \\
OpenHands  & gpt-5                 & 71.8\% & +14.0\% & +0.234 & *** \\
OpenHands  & claude-4-sonnet       & 70.4\% & +16.0\% & +0.264 & *** \\
SWE-agent  & claude-4-sonnet       & 66.6\% & +17.5\% & +0.297 & *** \\
OpenHands  & kimi-k2               & 65.4\% & +27.6\% & +0.388 & *** \\
CodeSweep  & kimi-k2               & 53.4\% & +89.2\% & +0.383 & *** \\
OpenHands  & claude-3.5-sonnet     & 53.0\% & +82.2\% & +0.393 & *** \\
OpenHands  & devstral-small        & 46.8\% & +44.3\% & +0.430 & *** \\
SWE-agent  & lm-32b                & 40.2\% &+111.9\% & +0.567 & *** \\
Skywork    & qwen-32b              & 38.1\% & +18.9\% & +0.173 & *** \\
SWE-agent  & claude-3.5-sonnet     & 33.6\% & +35.2\% & +0.364 & *** \\
SWE-agent  & gpt-4o                & 24.9\% & +54.4\% & +0.386 & *** \\
SWE-agent  & gpt-4                 & 22.6\% & +59.9\% & +0.506 & *** \\
SWE-agent  & claude-3-opus         & 15.1\% & +20.4\% & +0.506 & *** \\
\midrule
\multicolumn{6}{c}{\emph{Approach~B: fix task, vary agents (416 contested tasks, paired Wilcoxon)}} \\
\midrule
\multicolumn{2}{l|}{\textbf{Within-task aggregate}} & --- & \textbf{$-$10.0\%} & --- & \textbf{***} \\
\bottomrule
\end{tabular}

{*}{*}{*}\,$p<0.001$, {*}{*}\,$p<0.01$, {*}\,$p<0.05$.
\end{table}

\paragraph{Approach~B (fix task, vary agents).} To control task difficulty, we compare resolved and failed trajectories \emph{within each contested task}. For each of the 416 contested tasks, we compute the mean trajectory length for resolved and failed agents separately. Interestingly, the result reverses: resolved agents average 44.0 steps vs.\ 39.6 for failed agents (10.0\% longer). On 263 of 416 tasks (63\%), resolved trajectories are longer ($p = 1.9\times10^{-9}$, two-sided Wilcoxon signed-rank). On the same task, the agents that succeed take more steps than the agents that fail.

The two approaches give opposite answers, revealing that trajectory length simultaneously reflects task difficulty and agent capability. This ambiguity motivates RQ2b: we need behavioral dimensions that do not co-vary with task difficulty.

\subsubsection{RQ2b: Distinguishing Successful Strategies from Failed Ones.}
\label{sec:results-rq2b}

We examine how agents allocate their first 10~steps between reading and patching, and how this relates to resolution rate. 

\begin{figure}[!tp]
\centering
\includegraphics[width=.8\linewidth]{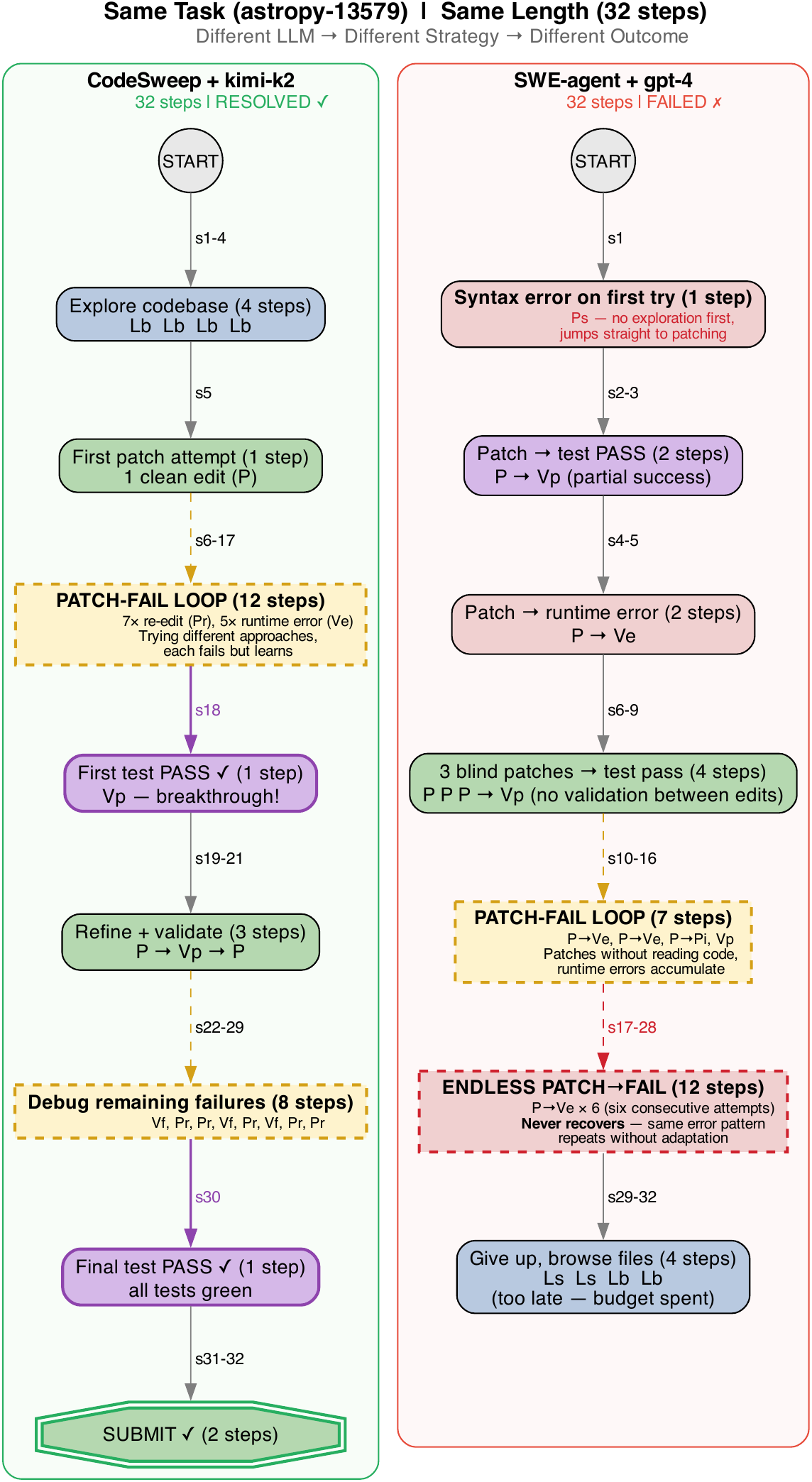}
\caption{Same task and length (32~steps), different agent, different outcome. \textbf{Left:} CodeSweep/kimi-k2 explores, struggles, breaks through (Vp), submits. \textbf{Right:} SWE-agent/gpt-4 skips exploration, enters an endless P$\to$Ve spiral, never recovers.}
\label{fig:rq2b-flowchart}
\end{figure}

\begin{figure*}[!tp]
\centering
\includegraphics[width=\textwidth]{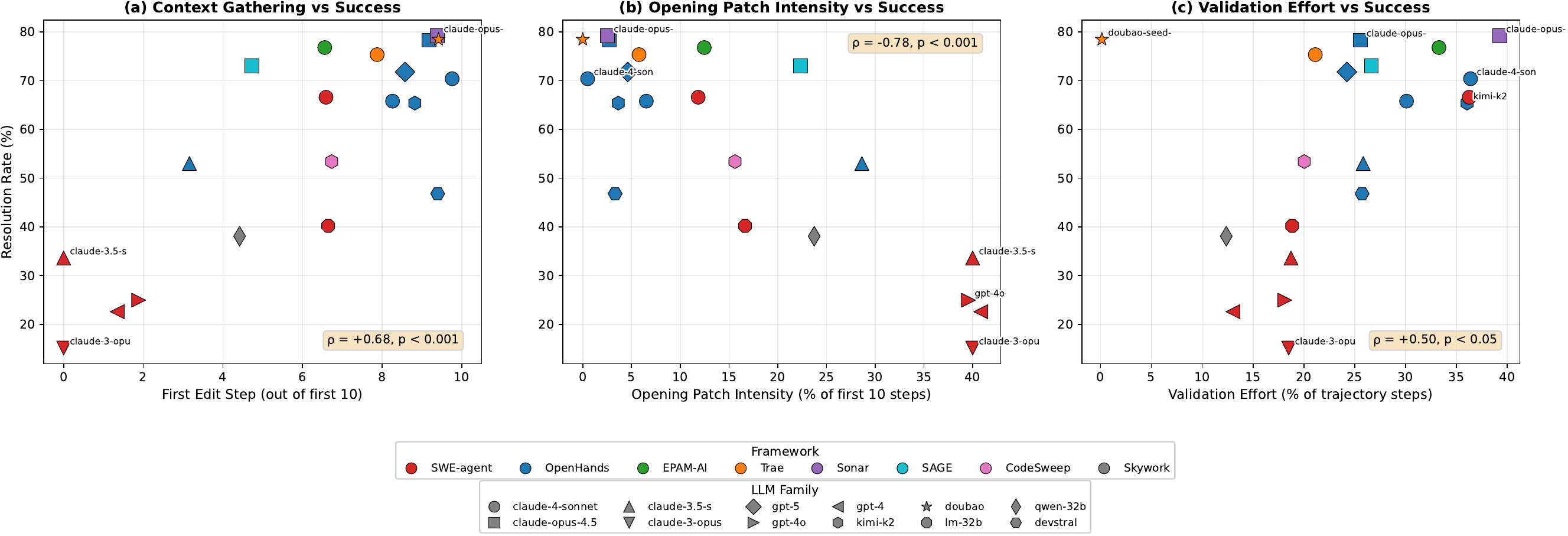}
\caption{Three structural dimensions of RQ2b across all agents. \textbf{(a)~Context gathering:} agents that delay their first edit succeed more ($\rho$ = $+$0.68, $p < 0.001$). \textbf{(b)~Opening patch intensity:} agents that spend more of their first 10~steps patching succeed less ($\rho$ = $-$0.78, $p < 0.001$). \textbf{(c)~Validation effort:} agents that spend more of their trajectory on validation succeed more ($\rho$ = $+$0.50, $p < 0.05$)}
\label{fig:rq2b-dimensions}
\end{figure*}

\paragraph{Context gathering and patch intensity.}
Figure~\ref{fig:rq2b-dimensions}a shows that agents delaying their first edit succeed more ($\rho = +0.68$, $p < 0.001$): agents at the top-right (e.g., claude-opus-4.5 at step~9.4) resolve most tasks, while those at the bottom-left (e.g., claude-3-opus at step~0) fail most. To illustrate this trend, Figure~\ref{fig:rq2b-flowchart} presents two agents with the same number of steps (32~each) on astropy-13579 but diverge entirely: CodeSweep/kimi-k2 explores the codebase first (4~Lb steps), struggles through a patch-fail loop, and eventually breaks through to a passing test; SWE-agent/gpt-4 patches immediately on step~1 and enters a P$\to$Ve spiral from which it never recovers. The inverse of context gathering is opening patch intensity (Figure~\ref{fig:rq2b-dimensions}b): agents that front-load patching in the first 10~steps succeed less ($\rho = -0.78$, $p < 0.001$). These results show how an agent spends its opening steps is strongly predictive of its overall success.

\paragraph{Validation effort.}
We examine how much of the trajectory agents spend on validation and how this relates to success. Validation effort (Figure~\ref{fig:rq2b-dimensions}c) correlates positively with resolution rate ($\rho = +0.50$, $p < 0.05$), ranging from 0.1\% (Trae/doubao) to 39.3\% (Sonar/claude-opus-4.5). One outlier stands out: Trae/doubao achieves 78\% resolution with near-zero validation, suggesting its LLM produces correct patches without test-driven iteration. At the other end, SWE-agent/gpt-4 validates only 13\% of steps and resolves 23\%. These results indicate that investing more trajectory steps in validation is associated with higher success, though sufficiently strong LLMs can bypass this need.

\begin{figure*}[t]
\centering
\includegraphics[width=\textwidth]{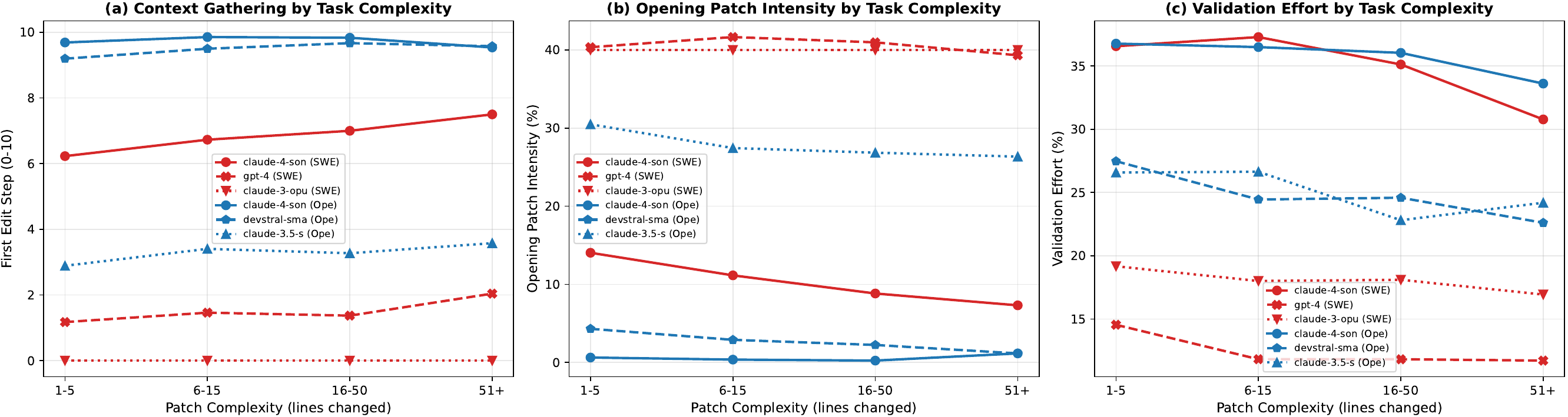}
\caption{Strategy stability across task complexity for 6~representative agents (3~SWE-agent in red, 3~OpenHands in blue). \textbf{(a)~Context gathering}, \textbf{(b)~opening patch intensity}, and \textbf{(c)~validation effort} are plotted against patch complexity (lines changed).}
\label{fig:rq2b-complexity}
\end{figure*}

\paragraph{Task-agnostic agent strategies.}
A natural question is whether agents adjust the three behavioral dimensions above (i.e., context gathering, patch intensity, and validation effort) based on task characteristics. Figure~\ref{fig:rq2b-complexity} tests this by plotting each dimension against patch complexity (lines changed) for six representative agents. For context gathering (Figure~\ref{fig:rq2b-complexity}a), the lines are flat: claude-3-opus edits at step~0 on every task regardless of complexity, while OpenHands/claude-4-sonnet waits until step~9.5 on every task. For opening patch intensity (Figure~\ref{fig:rq2b-complexity}b), the same pattern holds: agents that patch aggressively do so on simple and complex tasks alike. For validation effort (Figure~\ref{fig:rq2b-complexity}c), strong agents validate 35--37\% of steps across all complexity bins, while weak agents stay at 12--19\%. In all three figures, the vertical separation between agents (the level) is large, but the slope within each agent is near zero. This means the behavioral differences we observe are agent-determined rather than task-specific adaptations. Agents apply a fixed strategy regardless of what the task demands, suggesting that current agent designs lack mechanisms to assess task complexity and adjust their approach accordingly.

\begin{tcolorbox}[colback=gray!5, colframe=black, title=\textbf{Answering RQ2}]
Trajectory length is an ambiguous discriminator: the direction reverses depending on whether agent identity or task difficulty is controlled. Trajectory \emph{structure} discriminates consistently: agents that gather context before editing, avoid premature patching, and invest in validation succeed more. All three dimensions are agent-determined and stable across task complexity, indicating that agents employ fixed strategies that do not adapt to a given task.
\end{tcolorbox}

\paragraph{Implications.}
RQ2 has several practical implications. First, trajectory length should not be used as a standalone signal for failure detection and monitoring. Second, the structural dimensions we identify offer a more reliable alternative. The opening strategy (read-first vs patch-first) is observable within the first 10 steps and correlates with the outcome across all agents. A monitoring system that detects premature patching in the opening steps could flag runs unlikely to succeed before significant compute is spent. Finally, the strategy stability finding (Figure~\ref{fig:rq2b-complexity}) shows that agents use fixed strategies regardless of task complexity, suggesting that agent designers should focus on building adaptive strategies that adjust their approach to task requirements rather than adopting a one-size-fits-all workflow.

\subsection{RQ3: Does LLM capability or framework design drive agent success?}
\label{sec:results-rq3}

\begin{figure}[t]
    \centering
    \includegraphics[width=.75\columnwidth]{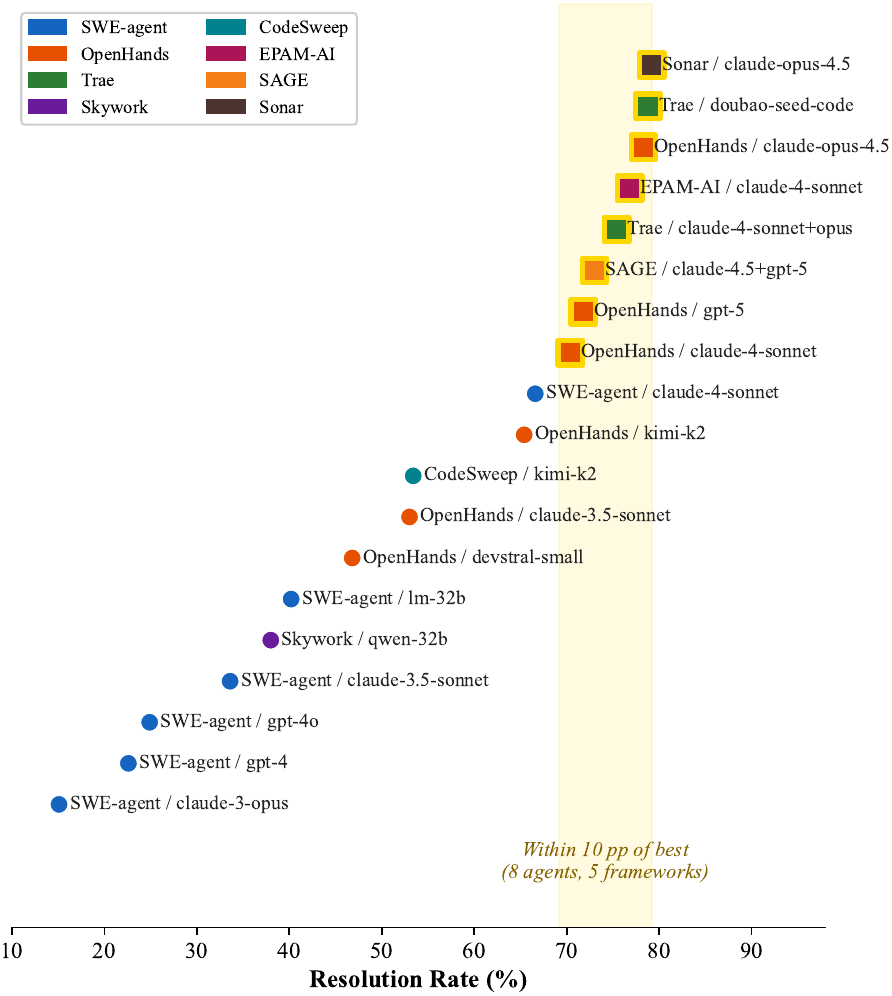}
    \caption{Resolution rate for 19 agents on SWE-bench Verified (500 tasks), colored by framework.}
    \label{fig:rq3-convergence}
\end{figure}

To answer this question, we first compare resolution rates across LLMs and frameworks in our dataset. Figure~\ref{fig:rq3-convergence} presents the resolution rate for all 19 agents. Within the same framework, swapping the LLM produces large shifts: 51.5~pp across 6 LLMs in SWE-agent, 31.5~pp across 6 in OpenHands. Holding the LLM constant, the framework effect is smaller: 3.8~pp for claude-4-sonnet, 19.4~pp for claude-3.5-sonnet. This gap narrows as LLMs improve (19.4~pp $\to$ 3.8~pp $\to$ 0.9~pp across three generations), and at the frontier, agents from 5 different frameworks converge to within 10~pp of the best.

\paragraph{Per-task agreement confirms LLM dominance.}

\begin{figure}[t]
    \centering
    \includegraphics[width=\columnwidth]{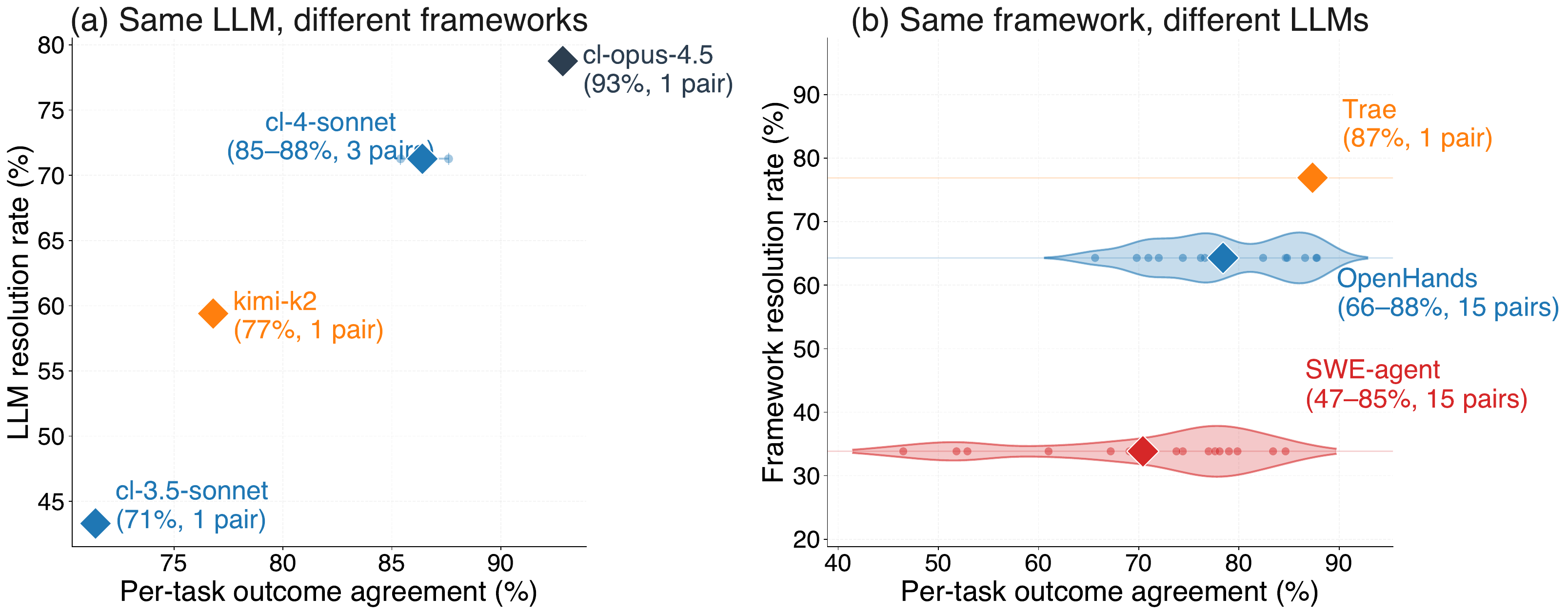}
    \caption{Per-task outcome agreement. (a)~Same LLM across frameworks: agreement increases with LLM capability. (b)~Same framework across LLMs: agreement is lower and more variable, shown as density distributions over all LLM pairs.}
    \label{fig:rq3-agreement}
\end{figure}

To move beyond aggregate resolution rates, we measure per-task outcome agreement: for each pair of agents sharing the same LLM (or the same framework), the fraction of 500 tasks on which both agents produce the same outcome (both resolve or both fail). Figure~\ref{fig:rq3-agreement} shows the results. When the LLM is held constant, and the framework varies (Figure~\ref{fig:rq3-agreement}a), agreement is high and increases with LLM capability: 71\% for claude-3.5-sonnet, 85--88\% for claude-4-sonnet (3~pairs), and 93\% for claude-opus-4.5. When the framework is held constant, and the LLM varies (Figure~\ref{fig:rq3-agreement}b), agreement drops substantially and spreads widely: 47--85\% across 15 LLM pairs in SWE-agent and 66--88\% in OpenHands. The density distributions in Figure~\ref{fig:rq3-agreement}b show that most LLM pairs agree on only 60--80\% of tasks, far below the 85--93\% seen when the same LLM is used across frameworks. This confirms that the LLM, not the framework, is the primary determinant of which tasks get solved.

\begin{figure*}[!tp]
    \centering
    \includegraphics[width=\textwidth]{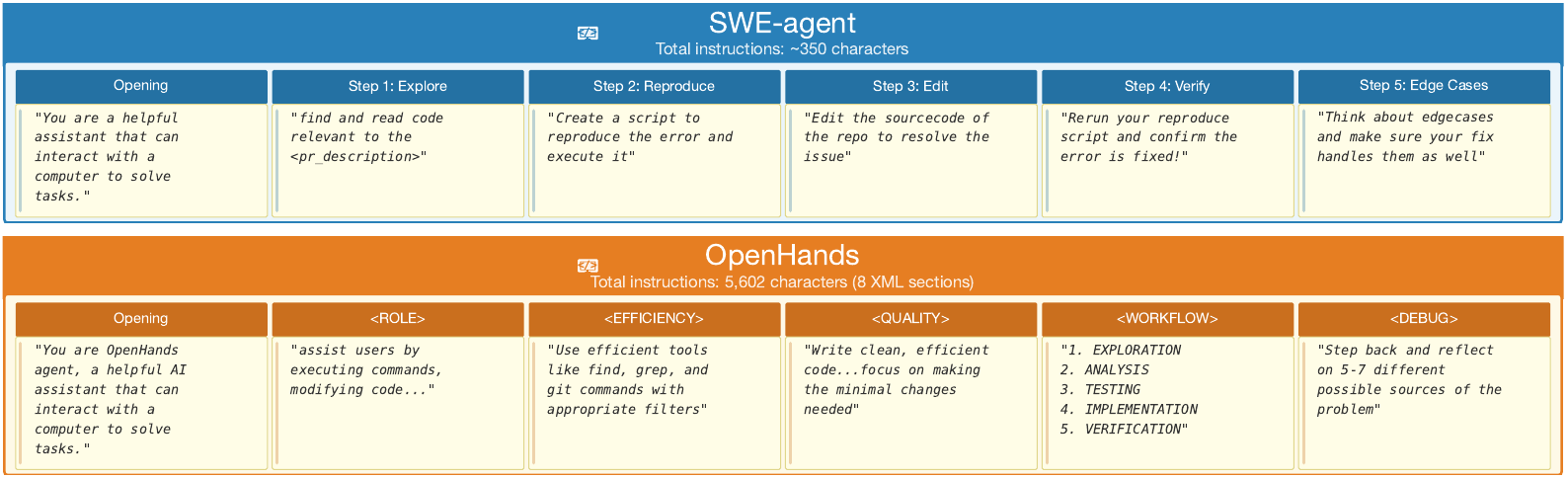}
    \caption{System prompt comparison for claude-4-sonnet on SWE-agent (top) vs.\ OpenHands (bottom). Quoted text extracted verbatim from trajectory files. Despite a 16$\times$ difference in instruction length (350 vs.\ 5,602 characters), the LLM produces near-identical core behavioral metrics on both frameworks.}
    \label{fig:prompt-comparison}
\end{figure*}

\paragraph{Prompt influence diminishes with LLM capability.}

We examine how much the framework's system prompt shapes agent behavior by comparing the same LLM on SWE-agent vs.\ OpenHands (Figure~\ref{fig:prompt-comparison}). These two prompts differ substantially: SWE-agent's is 350 characters, OpenHands' is 5,602. Each prompt does steer the agent toward different tool usage patterns (e.g., SWE-agent produces more reproduction scripts, OpenHands produces more search actions). However, the total amount of validation work is nearly identical across frameworks (18.2 vs.\ 20.0 actions per trajectory), indicating that the prompts change \emph{which tools} the agent uses but not \emph{how much work} it does. Crucially, this prompt influence is capability-dependent: for claude-3.5-sonnet, switching frameworks yields an additional 20~pp improvement in resolution, but for claude-4-sonnet, the same switch yields only a 4~pp improvement in resolution. Stronger LLMs develop their own strategy and are less susceptible to prompt-level steering.

\begin{tcolorbox}[colback=gray!5, colframe=black, title=\textbf{Answering RQ3}]
The LLM is the primary driver of both outcome and behavior; the framework's contribution shrinks with each LLM generation. Two agents sharing the same LLM agree on 85--93\% of tasks regardless of framework, whereas two agents sharing the same framework but different LLMs agree on only 47--88\% (Figure~\ref{fig:rq3-agreement}). The framework performance gap narrows from 19.4~pp to 3.8~pp to 0.9~pp across 3 successive Claude generations. Framework prompts do shape tool choices but not overall effort, and this shaping effect diminishes as LLMs grow stronger.
\end{tcolorbox}

\paragraph{Implications.} RQ3 shows that at the current frontier, investing in better LLM reasoning will yield larger returns than redesigning the orchestration scaffold. The convergence pattern reinforces this: as LLMs improve, the gaps in the framework shrink. Framework engineering is not irrelevant, but its marginal value diminishes with each generation of LLM.
For practitioners selecting an agent, this finding suggests that the choice of LLM matters more than the choice of framework: allocating budget toward a stronger model is likely to produce larger gains than switching scaffolds.
For researchers, this result implies that comparing two agents without controlling for the underlying LLM conflates model capability with system design. Future empirical studies should adopt factorial designs (LLM $\times$ framework) to isolate each factor's contribution.
Our results also challenge the assumption that more detailed system prompts lead to better agent performance. SWE-agent provides a minimal prompt (350 characters), while OpenHands prescribes an 8-phase workflow in 5,602 characters---a 16$\times$ difference. Yet for claude-4-sonnet, both produce near-identical core behavioral metrics and only a 4\% resolution difference (Figure~\ref{fig:prompt-comparison}). The additional prompt detail adds overhead (60 vs.\ 56 median steps) without improving outcomes. Prior work has shown that LLMs are highly sensitive to prompt formatting~\cite{sclar2024prompt} and that longer inputs can degrade reasoning performance~\cite{levy2024length}, though larger models exhibit enhanced robustness to prompt variations~\cite{zhuo2024prosa}. Our finding extends this to the agent setting: for strong LLMs, lean prompts may be preferable, as the model's own reasoning compensates for the lack of explicit guidance while verbose prompts add procedural overhead. The effect is capability-dependent, however: weaker LLMs (claude-3.5-sonnet) do benefit from richer prompts, gaining 19.4~pp from the same framework switch.

\section{Limitations and Threats to Validity}
\label{sec:threats}

\paragraph{Internal validity.}
Our study uses observational data, which may introduce confounds. While within-task comparisons control for task difficulty, other factors, such as tool API differences (e.g., edit vs.\ str\_replace) and framework prompting, may influence behavior. We mitigate this through cross-framework comparisons with shared LLMs and within-framework LLM variation, but cannot fully eliminate residual confounding. Our trajectory encoding relies on deterministic regex parsing, which may introduce minor classification errors (18/\numtrajectories failures). Additionally, agent trajectories were collected at different points in time, so differences in API versions, hardware, or SWE-bench infrastructure could affect results. We mitigate this by analyzing only final submitted trajectories under identical evaluation conditions (the SWE-bench harness).
Because we observe each agent on only a single run per task, we cannot measure the stochastic variance inherent in LLM-based agents. Prior work~\cite{brown2024monkeys} shows that repeated runs can change outcomes; our single-run design may therefore over- or under-attribute failures to deterministic causes.

\paragraph{Construct validity.}
Our 13-symbol trajectory encoding captures actions and outcomes but abstracts away finer-grained differences in reasoning and chain-of-thought quality. Metrics such as trajectory length and validation effort are proxies for strategy and may not fully capture underlying cognition. We mitigate this by combining multiple behavioral dimensions with qualitative analysis.
The classification of tasks into ``never-solved,'' ``contested,'' and ``always-solved'' is population-dependent: a different set of agents could shift these categories. Similarly, our \numtaskfeatures task features are computed from observable artifacts (patches, issues, metadata) and may miss latent difficulty factors such as conceptual reasoning complexity or implicit domain knowledge not reflected in the issue text.

\paragraph{External validity.}
Our results are based on 500 Python tasks across \numpythonframeworks open-source repositories from SWE-bench Verified and may not generalize to other programming languages, proprietary codebases, or development settings.

\paragraph{Conclusion validity.}
We use non-parametric statistical tests (Mann-Whitney $U$, Wilcoxon signed-rank) and effect sizes (Cliff's $\delta$) over a large dataset (\numtrajectories trajectories), supporting robust comparisons. However, some analyses involve small samples: the 12 never-solved simple-patch tasks (RQ1) and the 25 always-solved tasks limit statistical power for subgroup comparisons. We complement these with qualitative analysis to strengthen conclusions. For RQ3, the number of same-LLM and same-framework pairs is constrained by the available leaderboard submissions, which limits the generalizability of the agreement analysis to specific LLM--framework combinations rather than all possible pairings.

\section{Related Work}

Three lines of work study aspects of coding agent performance:

\paragraph{Task difficulty characterization.}
Prior work primarily defines task difficulty using patch-based metrics. Multi-SWE-bench~\cite{zan2025multiswebench} categorizes tasks by human-estimated resolution time, showing that harder tasks involve larger patches. SWE-bench Pro~\cite{swebenchpro2025} and Ganhotra~\cite{ganhotra2025cracking} similarly rely on patch complexity, while Large Language Monkeys~\cite{brown2024monkeys} highlights stochastic solvability without analyzing difficulty drivers. Although such metrics correlate with difficulty in our data ($\rho = -0.380$), they fail to explain edge cases: 12 tasks requiring minimal edits (single-file, $\leq$10 changes) are unsolved by all agents. Prior work does not investigate why these seemingly simple tasks remain unsolvable.

\paragraph{Behavioral analysis of coding agents.}
Existing studies examine agent behavior but lack proper controls or scale. Majgaonkar et al.~\cite{majgaonkar2025} report longer failed trajectories but do not control for task difficulty, leading to confounded conclusions. Bouzenia and Pradel~\cite{bouzenia2025understanding} identify repetitive loops and cascading errors, but analyze only 120 trajectories across three agents. Chen et al.~\cite{chen2026beyondfinalcode} propose a process-oriented error taxonomy but do not perform within-task paired comparisons. Chen et al.~\cite{chen2026rethinking} scale analysis across six LLMs but focus only on generated tests, finding that test-writing practices provide marginal utility rather than differentiating success from failure.

\paragraph{Decomposing agent performance.}

Other work compares agents at the architecture or LLM level. Martinez and Franch~\cite{martinez2025dissecting} and Ceka et al.~\cite{ceka2025understanding} treat agents as monolithic, conflating LLM and framework effects. More recent studies begin to vary LLMs: SWE-bench Pro~\cite{swebenchpro2025} categorizes failures using LLM summaries, SWE-Compass~\cite{swecompass2025} evaluates multiple LLMs across frameworks at the outcome level, and Ehsani et al.~\cite{ehsani2026where} analyze large-scale PRs without LLM-level breakdown. At the task level, DEI~\cite{zhang2024dei} shows that different agents resolve different sets of tasks and achieves 34.3\% resolution through ensembling open-source agents that individually score at most 27.3\%, demonstrating that task-level agreement across agents is low. However, DEI treats each agent as monolithic and does not decompose whether the disagreement stems from the LLM or the framework. We address this gap by measuring per-task outcome agreement separately for same-LLM/different-framework pairs and same-framework/different-LLM pairs, showing that LLM identity predicts task-level outcomes more than framework identity.

Across all three lines of work, no prior study has combined within-task behavioral analysis with cross-framework LLM decomposition at scale, which shows the novelty of our study.

\section{Conclusion}
\label{sec:conclusion}

This study presents a large-scale behavioral analysis of LLM-based coding agents, connecting task characteristics, agent architecture, and trajectory dynamics to understand how and why agents fail. We find that standard difficulty metrics leave edge cases unexplained where the bottleneck is architectural reasoning and domain knowledge rather than patch complexity.  Comparing success and failure trajectories under controlled paired designs shows that trajectory length is confounded by task difficulty, but structural properties of the trajectory, particularly how agents sequence reading, patching, and validation, distinguish success from failure consistently. Decomposing agent performance reveals that LLM capability drives both outcome and behavioral strategy, while framework design contributes surface-level variation that diminishes as models improve. Future work could extend this analysis to multilingual benchmarks, develop real-time monitoring tools that leverage behavioral signatures for early failure detection, and investigate whether the architectural reasoning gap can be addressed through retrieval-augmented or design-aware agent architectures.

\section*{Data Availability}
This paper includes supplementary artifacts available at~\cite {supplementar}.

\balance
\bibliographystyle{ACM-Reference-Format}
\bibliography{references}

\end{document}